\documentclass[12pt]{iopart}

\usepackage{iopams}  
\usepackage{graphicx}
\begin{document}

\title[SOC  in Metamaterials and Metasurfaces]{ Spin-Orbit Angular Momentum Conversion  in Metamaterials and  Metasurfaces}

\author{Graciana Puentes$^{1,2}$ }

\address{\emph{1-Departamento de Física, Facultad de Ciencias Exactas y Naturales, Universidad de Buenos Aires, Ciudad Universitaria, 1428 Buenos Aires, Argentina}\\

\emph{2-CONICET-Universidad de Buenos Aires, Instituto de Física de Buenos Aires (IFIBA), Ciudad Universitaria, 1428 Buenos Aires, Argentina}\\
%\emph{3- Department of Photonics Engineering, Technical University of Denmark, Ørsteds Plads, Building 343, DK-2800 Kgs. Lyngby, Denmark}
}
\ead{gpuentes@df.uba.ar}
\vspace{10pt}
%\begin{indented}
%\item[]August 2017
%\end{indented}

\begin{abstract}
In the last decades unprecedented progress in the manipulation of spin angular momentum (SAM) and orbital angular momentum (OAM) of light has been achieved, enabling a number of applications ranging from classical and quantum communication, to  optical microscopy and super-resolution imaging. Metasurfaces are artificially engineered 2D metamaterials with designed subwavelength-size building blocks, which allow precise control of optical fields with unparalleled flexibility and performance. The reduced dimensionality of  optical metasurfaces enables new physics and leads to functionalities and applications that are remarkably different from those achievable with bulk materials. In this review, we present an overview of the progress in optical metasurfaces for manipultation of SAM and  OAM of light, for applications in integrated spin-orbit conversion (SOC) devices.
\end{abstract}

\section{Introduction}
\label{sec1}
It is well known that, besides  linear momentum, electromagnetic (EM) waves possess angular momentum (AM) {\cite{review1, review2, review3,review4,review5}}, which can be decomposed into  spin angular momentum (SAM) and orbital angular momentum
(OAM). Circularly polarized waves carry SAM, quantized as $\pm \hbar$ per photon for left-hand circular (LHC) and right-hand circular (RHC) {polarization}, respectively. 
OAM describes the {azimuthal} angular dependence of photons and is quantized as $ l \hbar$ per photon, with $l$ any integer. Different $l$-values correspond to mutually orthogonal
OAM modes and the total numberof  OAM modes  is unbounded. OAM modes have been used to encode information and thus to enhance the channel transmission capacity, via OAM multiplexing and multicasting techniques %ref. 6,8,10,11,52,53,54,61,63 and 81 are not mentioned in the text, please check. and the order of ref.s citations are wrong, please reorder them.
{\cite{review1,review2}}.  Optical vortices  characterizing OAM beams also find a number of applications in super resolution imaging, optical tweezers, detection the rotation of particles at visible, THz and microwave regions \cite{review1, review2, review3, review4, review5, Veselago1968, Pendry2000, Pendry2006, Sououlis2010, Zheludev2010, Zhang2016, Yuetal2011, Ni2012. opex2018, Cao2014, Hui2015, Photonics2019, Science2013, Nanoscale20191, Science2011, PRB2018, NatMater2014, NanoLett2011, NanoLett2019, Nanoscale20192, SciRep2012}.

2D metamaterials, also known as metasurfaces {\cite{review2, review3,review4}}, is an emerging interdisciplinary field which promotes the of use  alternative approaches for {light engineering}  based on subwavelength-thick metasurfaces built upon metaatoms, with spatially varying compositions. They exhibit remarkable properties in {maneuvering}  light at a 2D interphase. Metasurfaces can not only achieve the functionalities of the 3D counterparts, such as negative refractive index and invisibility cloaking~{\cite{Pendry2000, Pendry2006}} but they can also resolve some of the current limits present in 3D metamaterials, such as {dielectric loss or  high resistive (ohmic) loss \cite{review1, review2, review3}}. In addition, metasurfaces can be easily fabricated via standard nanofabrication {approaches}, such as electron-beam lithography, {processes} readily available in the semiconductor industry.

In this review, we present some of the recent advances in 2D metamaterials and metasurfaces and their major applications in generation  and manipulation of SAM and OAM of light via spin-orbit-conversion (SOC). In Section \ref{sec2} we describe the basic building blocks of metasurfaces.  {In Section \ref{sec3}, we present a theoretical  introduction to the concepts of SAM, OAM and Total Angular Momentum (TAM) of light. In Section \ref{sec4}, we present the use of metasurfaces in the control of  SAM, OAM and TAM of light via q-plates and J-plates. In Section \ref{sec5}, we review current applications of metasurfaces as OAM multiplexing and multicasting techniques for high speed optical communication. Finally, in Section \ref{sec6} we outline the conclusions.}

\section{Metasurface Components}
\label{sec2}

Metasurfaces are 2D artificially  engineered materials made of subwavelength metaatoms which {permit} exotic optical and electromagnetic properties unavailable in natural compounds \cite{review1,review2,review3, Photonics2019, Science2013, Nanoscale2019, Science2011, PRB2018, NatMater2014, JPC2018, NanoLett2011, NanoLett2019, SciRep2012}. Due to the unique wave behaviors that result from {strongly} localized light-matter interaction, metasurfaces have the ability to overcome many tough problems faced by traditional materials, such as breaking of the diffraction limit, the generalized laws of refraction and reflection, as well as the localized enhancement of light absorption. 2D metamaterials exhibit enhanced light-interaction via plasmonic resonances 
 \cite{Ni2012, Kuester2003, Holloway2005}. When  a resonant external electric field is applied, electrons inside plasmonic materials are shifted from their steady-state positions resulting in a macroscopic polarization, which in turn generates an internal electric field to restore the electrons to their steady-states. Under harmonic excitation, collective oscillations can be generated. Depending on the size of plasmonic particles, they can be polarized either completely (small particles compared to wavelength) or only at the surface (larger particles compared to wavelength), resulting in simple dipolar resonances or multipole resonances, respectively. The resonant frequency depends on  size, shape, refractive index and surrounding media
 {\cite{review2,review3,review4,review5}}. On a metasurface, secondary waves reflected or transmitted through the nanoparticles acting as nanoantennas will gain different phase shifts and thus they can interfere and generate arbitrary wave fronts
  {\cite{Zhao2011,Pozar1997,Ni2012, Kuester2003, Holloway2005}}.

In order to increase the phase shift, the shape of the nanoantenna can be artificially engineered.  V-shaped nanoantennas
{\cite{Zhao2011,Pozar1997,Ni2012, Kuester2003, Holloway2005}}  can support two resonant modes: a symmetric mode and an {asymmetric} one. In coherent superposition of such resonance modes, a phase shift of 2$\pi$ can be successfully generated, for light whose polarization is perpendicular to that of the incidence. In order to mitigate the low efficiency associated with polarization conversion,  an alternative design can be employed incorporating a metallic ground plane separated from the top nanoantenna array by a thin dielectric layer 
{\cite{Zhao2011,Pozar1997,Ni2012, Kuester2003, Holloway2005}}. Light incident will induce antiparallel electric currents on the nanoantennas and ground plane, creating a gap resonance and extending the phase shift, ideally to 2$\pi$. Because plasmonic materials can provide strong optical resonances, they were {initially} chosen as the building blocks of metasurfaces 
\cite{Zhao2011,Pozar1997,Ni2012, Kuester2003, Holloway2005}. However, an important limitation of  plasmonic metasurfaces is energy dissipation into heat
\cite{Ni2012, Kuester2003, Holloway2005} or resistive loss.  Such resistive loss reduces the overall efficiency of plasmonic devices, thus precluding them from important applications. Additionally, due to the great capacity of  plasmonic structures  in  light confinement at the nanoscale, incident light fields can significantly  increase the local temperature, for this reason such plasmonic structures usually present a low  power threshold of deformation \cite{Zheludev2010}. Noble metals, such as gold and silver, are widely used as the building components for plasmonic nanostructures, {which can increase} the fabrication cost and can be incompatible with current semiconductor fabrication processes.

To overcome some of these limitations, alternative  metasurface designs  using only dielectric components have been proposed 
\cite{Zhao2011,Pozar1997,Ni2012, Kuester2003, Holloway2005}. Nanostructures composed of dielectrics with high refractive indices, such as silicon, germanium and tellurium, can manipulate light through Mie resonances \cite{Kuester2003,Holloway2005,Zhao2011,Pozar1997} 
and can achieve  both {electric and magnetic field} enhancement. A significant advantage of dielectric materials is that  they are associated with low resistive loss, due to small imaginary part of permittivity and can in principle {allow}  highly {efficient} devices. Moreover, fabrication of dielectric metasurfaces is compatible with state of the art semiconductor  processes. As a simple example to illustrate the resonance of dielectric nanoantennas, for sphere-shaped resonators the two lowest resonant modes excited are the electric and magnetic dipolar resonances \cite{Zhao2011,Pozar1997,Ni2012, Kuester2003, Holloway2005}. %plase confirm if it is  ref. citation.
By overlapping such lower order  resonances, it is possible to achieve a phase difference over the entire  2$\pi$-range. In~addition, when electrical and magnetic dipole moments have perpendicular orientations,  scattered light is unidirectional, which can be exploited to make reflection-free metasurfaces \cite{Zhao2011,Pozar1997,Ni2012, Kuester2003, Holloway2005}. %plase confirm if they are  ref. citations.
Aside from metallic metasurfaces based on plasmonic antennas, there is growing  interest in  nanoresonators exhibiting Fano resonances, arising from  interference between broad spectral lines  with  narrow  resonances \cite{Ni2012, Kuester2003, Holloway2005,Ryan2010,Caiazzo,Holloway2008}.
By designing antennas with broken  symmetry, weak coupling of plasmonic modes to free-space radiation modes can enable Fano resonances with narrow linewidths and high Q-factor. Fano resonances are also observed in pure dielectric systems through coupling of magnetic and electric dipolar modes 
\cite{Ryan2010,Caiazzo,Zhao2011,Pozar1997,Ni2012, Kuester2003, Holloway2005}. The Fano resonance is directional due to constructive interference. Moreover, the scattering direction is sensitive to frequency shift, which can be exploited in sensing and  switching  applications {\cite{Cao2014}}. 

\begin{figure}
\centering
%\begin{centering}
\includegraphics[width=0.45\textwidth]{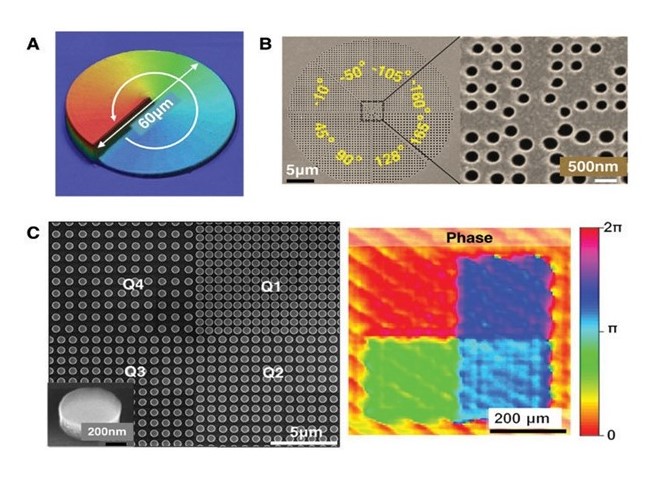}
\caption{Optical vortex beam generated by dynamic phase plates. (a) Optical profilometry image of a 60 $\mu$m diameter 3D femtosecond laser printed mircospiral phase plate (SPP) with continuous phase changing generates optical vortex with topological charge $l=5$ at 633 nm. (b) Nanowaveguide array with a phase modulation covers full range, generating optical vortex with topological charge $l=1$ at 532 nm. (c) Silicon nanodisks with high refractive index. Reprinted from Reference \cite{review1}.}

%\end{centering}
\label{fig1}
\end{figure}

Finally, there is an alternative approach for phase modulation in metasurfaces based on the geometric-phase (or Berry phase) elements to achieve tunable phase shifts in metasurfaces. It should be noted that typically this approach works only for circularly polarized light %plase confirm if they are  ref. citations.
\cite{Engheta,Papakpostasetal,Prosvirnin,Beth36,Woerdman92,Prosvirnin2009,Svirko}. The building blocks of geometric phase metasurfaces can be described as polarization wave-plates with  spatially varying orientations of fast axes (see Figure \ref{fig1}A--C). Space-variant polarization {conversion} can be easily described using  Jones formalism for polarized light, obtaining an expression for the {transformation} matrix $T(x,y)$ {\cite{review1}}:
\begin{equation}
T(x,y)=R(\theta(x,y)) J(\phi)   R(\theta(x,y))^{-1},
\end{equation}
where $J(\phi)$ represents the Jones matrix of  a retarder plate with retardation $\phi$, $R$ is a rotation matrix and $\theta$ is the orientation of the fast axis of the rotation plate. As will be described in the following Sections, it is apparent that by transmission on a geometric-phase variant metasurface RHC or LHC light will be converted into the opposite circular polarization and will {acquire} an additional spatially dependent geometric phase of the form 2$\theta(x,y)$. By rotating the spatially varying waveplate from 0  to 180 degrees, the induced phase shift can be continuously tuned from 0 to 2$\pi$. This phase shift is purely geometric, therefore it has a broad bandwidth. Its performance is limited only by  the scattering efficiency of the nanoantennas. Moreover, geometric-phase elements can be resonantly excited to increase the scattering efficiency. The spatially varying phase shift  $\theta(x,y)$ can be {tailored} in order to produce controlled OAM vortex states. This transformation is the called Spin to OAM Conversion (SOC), as described in detail in the Section \ref{sec2}. Figure \ref{fig1} depicts different approaches to OAM generation in 2D Metamaterials and Metasurfaces.

\section{SAM, OAM and TAM of Light }
\label{sec3}
\vspace{-6pt}

\subsection{SAM of Light}

The SAM and OAM of light are separately observable properties in optics.  For paraxial monochromatic beams, a spin-orbital angular momentum decomposition is straightforward. This~unique feature explains in part the recent {unrivaled} progress in photonic  SAM-OAM  conversion in metasurfaces and 2D  {metamaterials}, which motivates this review. {At the same time, fundamental difficulties in quantum
electrodynamics and field theory  result in a number of subtleties for the spin and orbital
AM description in generic non-paraxial or non-monochromatic fields \cite{Bliokh2015, Bliokh2014}}. 

{Within the unified theory of angular momentum of light, based on canonical momentum and spin densities developed in Reference \cite{Bliokh2015}}, the spin angular {momentum} (SAM)  is associated with the polarization of light, so that right-hand circular (RHC) and left-hand circular (LHC) {polarization} of a paraxial beam correspond to  positive and negative {helicity} $ \sigma  = \pm 1$  of  a photon. If the mean momentum of the beam (in units of $\hbar$  per photon) can be associated with its
mean wave vector $\langle \textbf{k} \rangle$, then such beam carries the corresponding SAM $\langle S \rangle = \sigma \langle \bf{k} \rangle/| k| $, 
where the helicity parameter $\sigma$ is equivalent to the degree of circular polarization in the Jones formalism. 

{More specifically}, a plane wave is an idealized entity which extends to infinity. Such plane wave cannot carry  extrinsic OAM  (analogue to the mechanical counter part $\bf{L}={r} \times {p}$),
because its position $\bf{r}$ is undefined. On the other hand, a circularly-polarized electromagnetic plane wave can carry SAM. {In the canonical momentum representation, the vector describing the electric field, for a circularly polarized plane-wave propagating along the \emph{z}-direction can be written as \cite{Bliokh2015}}:

\begin{equation}
\bf{E} \propto \frac{\hat{x} + i \sigma \hat{y}}{2} \exp(i|k|z), 
\end{equation}
where $\hat{x} ,\hat{y} ,\hat{z}$ are unit vectors and the helicity parameter $\sigma  = \pm 1$ corresponds to the LHC and RHC polarizations, respectively. The wave number $|k|$, results from the dispersion relation for a plane wave, that is, $ |k| = \omega / c$. The electric field described in Equation (2) represents the eigen-mode of the \emph{z}-component of the spin-1 matrix operators with eigenvalue $\sigma$, of the form $\hat{S}_{z} \bf{E} = $ $\sigma$ ${\bf{E}}$ {\cite{Bliokh2015}}. {Where the spin-1 operators (generators of the SO(3) vector rotations)  are given by {\cite{Bliokh2015}}}:

\begin{equation}
\hat{S}_{x}=-i\left( \begin{array}{ccc}
0 & 0 & 0 \\
0 & 0 & 1 \\ 
0 & -1 & 0 \end{array} \right), \hspace{3mm} \hat{S}_{y}=-i\left( \begin{array}{ccc}
0 & 0 & -1 \\
0 & 0 & 0 \\
1 & 0 & 0 \end{array} \right), \hspace{3mm} \hat{S}_{z}=-i\left( \begin{array}{ccc}
0 & 1 & 0 \\
-1 & 0 & 0 \\
0 & 0 & 0 \end{array} \right). 
\end{equation}

Therefore, the plane wave carries SAM density $\langle S  \rangle = \sigma \frac{\langle \bf{k} \rangle}{|\bf{k}|}$, defined as the local expectation value of the spin operator with the electric field (Equation (2)) \cite{Bliokh2015}.

\subsection{OAM of Light}

\begin{figure}
\centering
%\begin{centering}
\includegraphics[width=1\textwidth]{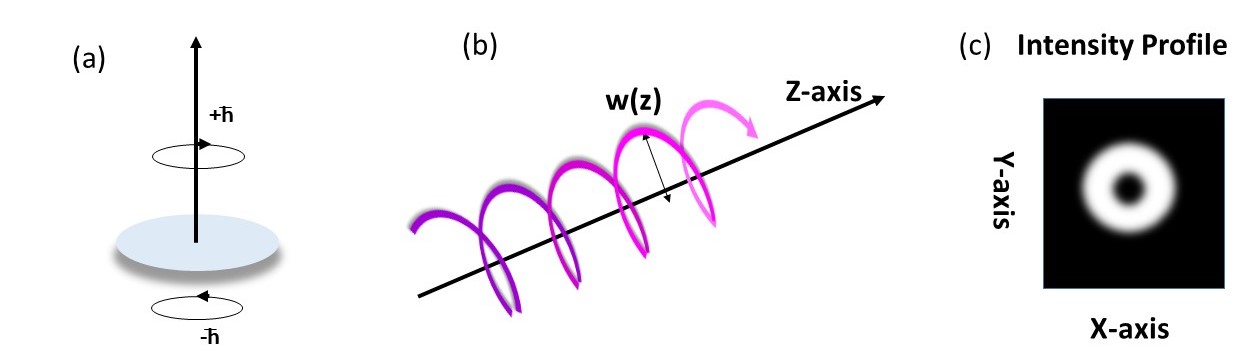}
\caption{(\textbf{a}) A suspended $\lambda / 2$ birefringent plate undergoes
a torque in transforming right hand ciruclar (RHC) into left hand circular (LHC)
polarized light \cite{Woerdman92}. (\textbf{b}) Helical curve representing the Poynting vector of
a linearly polarized Laguerre-Gaussian mode of radius $w(z)$. (\textbf{c}) Intensity profile of Laguerre-Gaussian beam on the transverse plane.}
%\end{centering}
\label{fig2}
\end{figure}

The first demonstration of the mechanical torque created by the transfer of angular momentum of a circularly polarized light beam to a {birefringent} plate was performed by Beth in 1936 \cite{Beth36}. In~this experiment a half wave-plate was suspended by a fine quartz fiber (Figure \ref{fig2}a). {A beam of circularly polarized light} by a fixed quarter-wave plate passed through the plate which {transform} RHC polarization, with spin component $-\hbar$, into LHC polarization, with spin component $+\hbar$, with~a net SAM transfer of $2 \hbar$ per photon to the {birefringent} plate. {Beth measured torque} agreed in sign and modulus with the quantum and classical predictions, his experiment is usually referred to as the measurement of SAM of the photon.

In Reference \cite{Woerdman92}, it was demonstrated that Laguerre-Gaussian modes, with azimuthal angular dependence $\exp{[-il\phi]}$, are  {eigen}-modes of the momentum operator $L_{z}$ and carry an orbital angular {momentum} $l \hbar$ per photon. A convenient representation of a linearly polarized TEM$_{plq}$ laser mode can be obtained, in the Lorentz gauge, using the vector potential \cite{Woerdman92}:

\begin{equation}
A=u(x,y,z)\exp{[-ikz]}\hat{x},
\end{equation}
where $\hat{x}$ is the unit vector in \emph{x}-dir and $u(r,\phi,z)$ is the complex {scalar} field amplitude satisfying the paraxial wave equation. In the paraxial approximation, second derivatives and the products of first derivatives of the electro-magnetic field are ignored and $du / dz$ is taken to be small compared to $ku$. The solutions for the cylindrically symmetric case $u_{r,\phi,z}$, describing the Laguerre-Gauss beam, are of the form \cite{Woerdman92}:
\begin{eqnarray}
u_{r,\phi,z}&=& \frac{C}{\sqrt{1+\frac{z^2}{z_{R}^2}}}[\frac{r \sqrt{2}}{w(z)}]^l L_{p}^{l} [\frac{2r^2}{w(z)^2}]\\ \nonumber
                  &   &  \times\exp{[\frac{-r^2}{w(z)^2}]}\exp{[\frac{_ikr^2z}{2(z^2+z_{R}^2}]} \exp{[-il\phi]}\\ \nonumber
                    &  &  \times \exp{[i(2p+l+1)\tan^{-1}[\frac{z}{z_{R}}]]},
\end{eqnarray}
where $z_{R}$ is the Rayleigh range, $w(z)$ is the radius of the beam, $L_{p} ^{l}$ is the associated Laguerre polynomial, $C$ is a constant and the beam waist is taken at $z=0$. Within this description, the time avarage of the real part of the {Poynting} vector $\epsilon_0 E \times B$, which is the linear momentum density, results in:

\begin{equation}
\frac{\epsilon_0}{2} (E ^{*} \times B + E \times B^{*}) = i \omega  \frac{\epsilon_0}{2} (u^{* }\nabla u - u \nabla u ^{*}) + \omega k \epsilon_0 |u|^2 \hat{z},
\end{equation}
where $z$ is the unit vector in \emph{z}-direction. When applying to a Laguerre-Gaussian distribution given in Equation (4), the linear momentum density takes the form:

\begin{equation}
P=\frac{1}{c} [\frac{rz}{z^2+ z_{R} ^2}|u|^2 \hat{r} + \frac{1}{kr}|u|^2 \hat{\phi} + |u|^2 \hat{z}],
\end{equation}
where $\hat{r}, \hat{\phi}$ are unit vectors. It may be seen that the Poynting vector ($c^2 P$) spirals along the direction of propagation along the beam (Figure \ref{fig2}b). The $r$ component relates to the spatial spread of the beam, the $z$ component relates to the linear momentum and  the $\phi$ component gives rise to the OAM.

\subsection{OAM Modes and Topological Charge}

Optical vortices  characterizing  beams of photons propagating with a phase  singularity of the form $\exp[i l \phi]$, with $l$ is an integer number, have a  topological structure on its wavefront characterized by a  topological charge $l$, arising from its helical spatial wavefront which spirals around the phase singularity, as depicted in Figure \ref{fig2}b. The topological structure in the wavefront is not only limited to EM waves but can also be found in acoustic waves, electrons and neutrons  \cite{review3}. At the singularity, the~phase is not defined and the polarization and amplitude vanish altogether, resulting in a dark centre within the wavepacket, as depicted in Figure \ref{fig2}c. Such singularity can be considered  a screw dislocation. It is remarkable that optical dislocations have not only been observed in spacial Laguerre-Gauss modes but they were also predicted and observed in laser scattering speckle fields, resulting from the interfering of multiple plane waves. Therefore, optical dislocations are currently considered a universal phenomenon. A high degree of control over the different approaches for generation and manipulation of vortex beams carrying OAM has been achieved  in the last couple of decades, at~an unparalleled rate. Similarly impressive are the potential avenues of applications ranging from  optical microscopy, {microengineering}, to classical and quantum communications, among other exciting research directions~\cite{review1, review2, review3, review4, review5}.

\subsection{TAM of Light}

As described in previous subsections, beams with azimuthal phase profile $\exp{[i l \phi]}$, with $l$ an integer number, carry OAM. Since $l$ is any integer, possible values of OAM are unbounded,  of the form $l \hbar$ per photon. On the other hand, circularly polarized light carries SAM, taking only two possible values of the form $\sigma \hbar$ per photon, with $\sigma = \pm 1$. A paraxial optical beam can carry both OAM and SAM, quantified in the Total Angular Momentum (TAM) of the form $J=(l + \sigma) \hbar$ per photon.

{Geometric phase elements} \cite{Karimi Nature2014}, also known as $q$-plates, can couple SAM and OAM by converting a given state of circular polarization into its opposite, while imprinting an azimuthal phase profile with a {fixed} topological charge. On the other hand, generalized metasurfaces \cite{DelvinScience} can transform arbitrary input SAM states (i.e., elliptical polarization)  into arbitrary output states of opposite TAM. Such devices are also known as $J$-plates. In the next {Section} we provide an overview of the transformations that can be achieved with $q$-plates and $J$-plates respectively, using tailored metasurfaces.

\section{SAM, OAM and TAM Control in 2D Metamaterials} 
\label{sec4}
\vspace{-6pt}

\subsection{SAM Control in Metasurfaces}

In the same manner as conventional wave plates, metasurfaces {warrant}  polarization conversion through {maneuvering} of two eigenmodes of light corresponding to orthogonal {polarization}. To~design a metasurface for polarization conversion, one should know the Jones vector of the incident field and the Jones vector of the desired output fields. With such knowledge, a Jones matrix  $J(x,y)$ for each spatial point $(x,y)$ on the metasurface plane linking the incident and output waves can be determined. The calculated Jones matrix can be achieved by engineering  proper nanoantenna patterns. Up to now,  conversion between a linear polarization and a circular polarization, between different linear polarization states or between  opposite circular polarization states {have} been reported using metasurfaces {\cite{DelvinScience, Karimi Nature2014, Capasso SciRep}}. Geometrical phases can be generated by engineering  the metasurface with identical nanoantennas of spatially varying orientations, as described in Section \ref{sec1}. One major limitation of  geometric-phase metasurfaces is that they work only for circularly polarized  input light, therefore~the geometric phase must be carefully matched to the propagation phase to achieve arbitrary control of polarization states. This is typically achieved via hybrid patterns of nanoantennas.

\subsection{OAM Control in Metasurfaces}

Polarization-control in metasurfaces which {permit} {maneuvering} of the propagation phase and the geometrical phase can also be utilized  in the generation of optical vortex beams carrying net topological charge ($l$) or OAM of light %plase confirm if they are  ref. citations.
 \cite{Papakpostasetal, Beth36, Woerdman92, Prosvirnin2009, Svirko, Aieta,  Franke, Yao, Fang, Forbes, Willner, ref33, Padgett, Wang, Zhu}.  One of the most distinct feature of such OAM beams is their helixed wave front, a direct consequence of the dependence of  phase on the azimuthal angle. The order of OAM is labeled by an integer $l$,  given by  the number of the twists a wave-front contains (per unit wavelength). OAM has been widely recognized as an unbounded degree of freedom of light, which~can be utilized in high-speed free-space optical communication systems \cite{Ren,Franke,Yao,Fang,Forbes,Willner,ref33,Padgett,Wang,Zhu} %plase confirm if they are  ref. citations.
since beams with different orders of OAM  are orthogonal they do not interfere with each other.
Traditional methods for generation of  OAM beam include spatial light modulators (SLM), holograms,  laser-mode conversion and spiral phase plates (SPP) (Figure \ref{fig1}). On the other hand, metasurfaces can create helical wave fronts by arranging nanoantennas with linearly increasing (or decreasing) phase shifts along the azimuthal direction \cite{Zhao,Durnin,ZhuWang,Li,Paterson,Litvin,Hakola,Ni2012}. 
As a result, a  metasurface can  add an optical vortex to the incident light wave front, thus converting SAM into OAM,  this transformation is also termed spin-to-orbit conversion (SOC)~\cite{Meinzer2000,Zhao2000,DelvinScience, Karimi Nature2014,Capasso SciRep}. 
Normally the SOC permits only the conversion of LHC and RHC polarization into states with opposite OAM, due to conservation of total angular momentum (TAM). By providing an additional phase shift in the azimuthal direction, a metasurface can convert circular polarizations into states with independent values of OAM. In recent years, conversion of light with arbitray elliptical polarization states into orthogonal OAM vortex states was demonstrated \cite{Zhao2000,Du2000}.

\subsection{TAM Control in Metasurfaces}

\begin{figure}
\centering
%\begin{centering}
\includegraphics[width=0.8\textwidth]{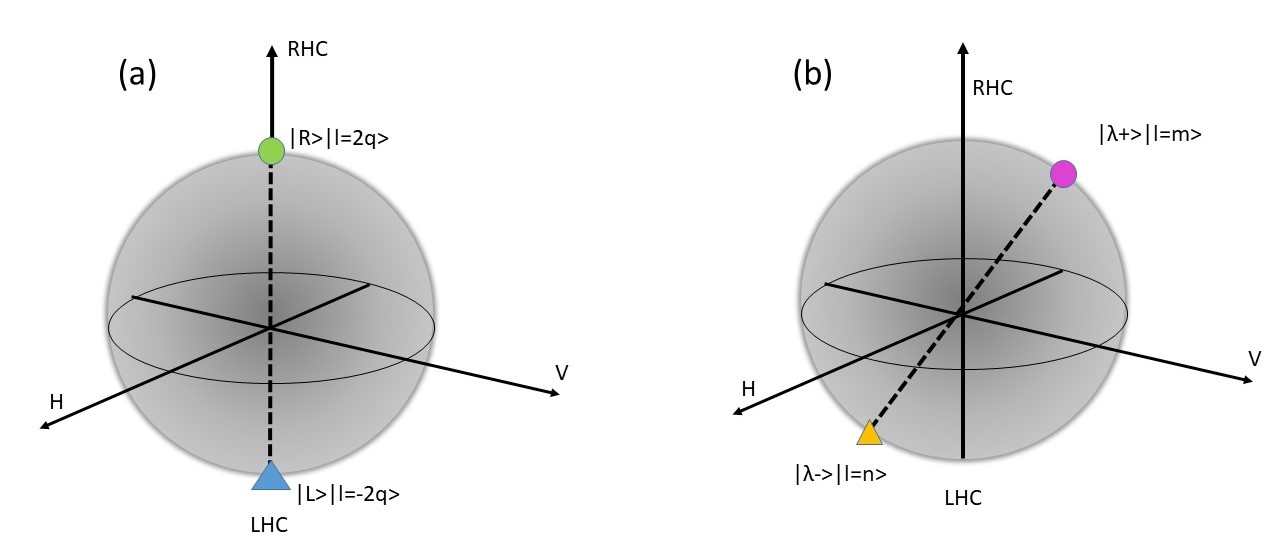}
\caption{(\textbf{a}) Representation in Poincare sphere of spin-orbit-conversion (SOC) via $q$-plates. A state of circular polarization located in the poles of the Poincare sphere is mapped into its opposite state of circular polarization, while imprinting a fixed {azimuthal} phase $\pm 2q \phi$. {(\textbf{b})} Representation in Poincare sphere of spin-orbit {conversion} (SOC) via $J$-plates. Arbitrary states of elliptic polarization $ | \lambda \pm \rangle$ are mapped into its opposite state of  elliptic polarization, while imprinting a tunable {azimuthal} phase~$ (n,m)\phi$. }
%\end{centering}
\label{fig3}
\end{figure}

Phase elements with spatially varying orientations can introduce a controlled geometric phase and provide for a connection between polarization (SAM) and phase (OAM). These devices are typically made of periodic elements referred to as $q$-plates \cite{Beth36}. The exact transformation performed by $q$-plates can be expressed as {\cite{DelvinScience, Mueller2017}}:
\begin{eqnarray}
|L \rangle& \rightarrow & \exp[i2q \phi ]| R \rangle\\ \nonumber
|R \rangle& \rightarrow & \exp[-i2q\phi] | L \rangle,
\end{eqnarray}
where LHC and RHC polarization are converted to its opposite spin state, with an OAM charge of $\pm 2q \hbar$ per photon (Figure \ref{fig3}a). This transformation is usually referred to as spin-orbit-conversion (SOC). As described in the previous section, typically the only elements that vary the angle spatially are those which perform the rotation (i.e., $\theta(x,y)$). For this reason the OAM output states are constrained to conjugate values ($\pm 2 q \hbar$). Moreover, due to the symmetry of the device, the SOC operation performed by $q$-plates is limited to SAM states of circular polarization (Figure \ref{fig3}a). In order to {perform} SOC operations in arbitrary SAM states, such as elliptic polarization, a more general device is required. This arbitrary mapping can be performed by  $J$-plates, as described below.

A $J$-plate has the {ability} to map arbitrary input SAM states, that is, not limited to RHC or LHC, into two arbitrary TAM states (represented by the variable $J$). A $J$-plate can be realized employing any medium that enables  {birefringence}, absolute phase shift and retarder orientation angles to vary spatially. In other words, in addition to the spatial orientation of the fast axes of rotation plates, the~relative phase shift between orthogonal spins should also vary spatially ($\phi=\phi(x,y)$). In~Reference~\cite{DelvinScience}, a $J$-plate was realized using metasurfaces. The exact transformation performed by the $J$-plate can be expressed as (Figure \ref{fig3}b):
\begin{eqnarray}
|\lambda+ \rangle& \rightarrow & \exp[im \phi ]|( \lambda +)^{*} \rangle\\ \nonumber
|\lambda - \rangle& \rightarrow & \exp[in\phi] | (\lambda -)^{*}\rangle,
\end{eqnarray}
where $|\lambda \pm \rangle$ are arbitrary elliptical polarization expressed as \cite{DelvinScience}:

\begin{equation}
|\lambda + \rangle=[\cos(\chi) , e^{i \delta} \sin(\chi)]^{T}, \hspace{1cm} |\lambda - \rangle=[-\sin(\chi) , e^{i \delta} \cos(\chi)]^{T},
\end{equation}
where the parameters ($\chi, \delta$) determine the polarizations state. Implementing the SOC operation reduces to determining the actual Jones matrix $J(\phi)$ that transforms $J(\phi)|\lambda+ \rangle=e^{im\phi}|(\lambda +)^{*} \rangle$ and  $J(\phi)|\lambda - \rangle=e^{in\phi}|(\lambda - )^{*} \rangle$. It can be demonstrated that the required spatially varying Jones matrix has the form:

\begin{equation}
J(\phi)=e^{i \delta} 
\left[ \begin{array} {cc}
 e^{i \delta} ( e^{i m \phi} \cos(\chi)^2 +e^{i n \phi} \sin(\chi)^2) &    \frac{\sin(2\chi)}{2}(e^{im\phi}- e^{in\phi})\\ \nonumber
 \frac{\sin(2\chi)}{2}(e^{-im\phi}- e^{-in\phi})                               &  e^{-i \delta} ( e^{i m \phi} \sin(\chi)^2 +e^{i n \phi} \cos(\chi)^2)\
\end{array}
\right ].
\end{equation}

We stress that no traditional phase plate can provide for the required control and sub-wavelength spatial variations in phase shift, {birefringence} and orientation. Sub-wavelength space metasurfaces on the other hand, can {allow} such control.

\subsection{Alternative Approaches for Spin-Orbit Conversion (SOC)}

\begin{figure}
\centering
%\begin{centering}
\includegraphics[width=0.7\textwidth]{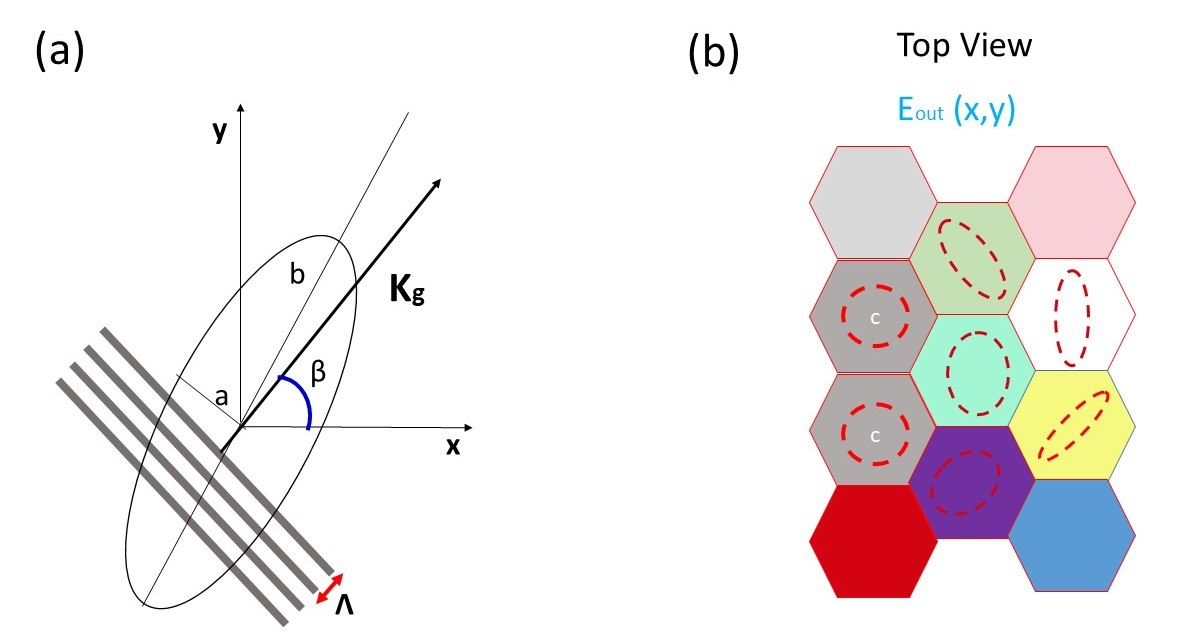}
\caption{(\textbf{a}) Geometrical parameters of polarization ellipse of light transmitted through a subwavelength varying grating  (SVG)  \cite{Bomzon2001, Kleiner2001}. (\textbf{b}) (Top view) Metasurfaces composed of elliptical nanoposts. The polarization and {azimuthal} phase of a {normally} incident optical wave $E_{in}(x,y)$ is modified at each pixel according to pixel design, resulting in a output optical wave $E_{out}(x,y)$ with controllable polarization and phase \cite{Faraon2015}.} 
%\end{centering}
\label{fig4}
\end{figure}

Controllable geometric-phases  can also be introduced via space-variant polarization-state {engineering} using subwavelength metal stripe space-variant gratings, as proposed by Bomzon {et al.}, in 2001 \cite{Bomzon2001, Kleiner2001}. This realization is based on a  subwavelength metal stripe space-variant grating (SVG) designed for converting circularly polarized light into an azimuthally polarized beam. SVGs are described by a vector:

\begin{equation}
%K_{g} (x,y) =2 \pi/ \Lambda (x,y)
 K_{g}(x,y)=2 \pi/ \Lambda(x,y)[\cos(\beta(x,y)) \hat{x} + \sin(\beta(x,y)) \hat{y}] ,
\end{equation}
where $\Lambda$ is the period of the grating and $\beta(x,y)$ is a spatially varying vector perpendicular to the metal stripes. The polarization of the transmitted beam through the grating can be expressed as:

\begin{equation}
J(x,y)=R[\beta(x,y)]J(\Lambda)R^{-1}[\beta(x,y)],
\end{equation}
where $J(\Lambda)$ is the Jones matrix representing the SVG of period $\Lambda$ for $\beta=0$,  and $R(\beta(x,y))$  is a~2~$\times$ 2 spatially varying rotation matrix (Figure \ref{fig4}a). $J(\Lambda)$, typically a diagonal matrix because the eigen-polarizations of the grating are oriented parallel and perpendicular the grating vector. Similarity between Equations (1) and (12) reveal that both the metasurface  and the SVG approach  produce a transformation described by the same Jones matrix.

In addition, successful alternative approaches for SOC using metasurfaces  based on high-contrast dielectric elliptical nanoposts have been recently  implemented \cite{Faraon2015}. The metasurface  is composed of a single-layer array of amorphous silicon elliptical posts with different sizes and orientations, resting on a fused-silica substrate. The elliptical posts are placed at the {centers} of hexagonal unit cells or pixels (Figure \ref{fig4}b). In brief, each post can be considered as a waveguide that is truncated on both sides and operates as a low-quality-factor Fabry–Perot resonator. The elliptical cross-section of the waveguide leads to different effective refractive indices of the waveguide modes polarized along the two ellipse diameters. As a result, each of the posts imposes a polarization-dependent phase shift on the transmitted  light field $E_{out}(x,y)$ and modifies both its polarization and phase. This~implementation provides complete control of polarization and phase with subwavelength spatial resolution and an experimentally measured efficiency ranging 72\% to 97\% from thus significantly improving transmission efficiency upon previous realizations \cite{Karimi Nature2014}. Such complete control {permits} the realization of most free-space transmissive optical elements such as lenses, phase plates, wave plates, polarizers, beamsplitters, as well as polarization-switchable phase holograms or arbitrary vector beam generators using the same metamaterial platform. 

\section{Applications in OAM Multiplexing and Multicasting {Approaches}}
\label{sec5}
Metasurfaces can contribute to a new class of chip-scale devices, which are expected to revolutionize nanophotonic and optoelectronic circuitry through smart integration of multiple functions in metallic, dielectric or semiconductor building blocks. Their 2D nature is a {significant} advantage for wafer-scale processing and integration, therefore among artificially engineered materials, metasurfaces hold great promise to excell in technologically relevant applications. Here we will focus on applications of metasurfaces in  integrated OAM multiplexing and multicasting devices, for~high-speed data transmission \cite{review1,Puentes1,Puentes2,Puentes3,Puentes4}.

OAM division multiplexing is an experimental {method} for increasing the transmission capacity of electromagnetic signals using the OAM of the electromagnetic waves to distinguish between the different orthogonal channels {\cite{ZhaoACSPhoton2018}}. It is the analogue of wavelength division multiplexing (WDM), time division multiplexing (TDM) or polarization division multiplexing (PDM) \cite{Jahani2000, Capasso SciRep},  %plase confirm if it is  ref. citation.
in the subspace spanned by OAM of light. While SAM) or polarization multiplexing offers only two orthogonal states corresponding to the two states of circular polarization, OAM multiplexing can  access a potentially unbounded set of states and as such can offer an infinite  number of channels for multiplexage. Although OAM multiplexing promises very significant improvements in bandwidth (2.5 Tbit/s transmission rates in MIMO systems have been reported), it is still an experimental {approach} and~has so far only been demonstrated in the laboratory, over relatively small distances of a few Kms over OAM maintaining fibers. In addition to the reduced transmission distances, one of the main limitations for {scalable} OAM multiplexage is the extreme bulkiness of the optical components required both for OAM generation and for OAM detection.

\begin{figure}
\centering
%\begin{centering}
\includegraphics[width=0.7\textwidth]{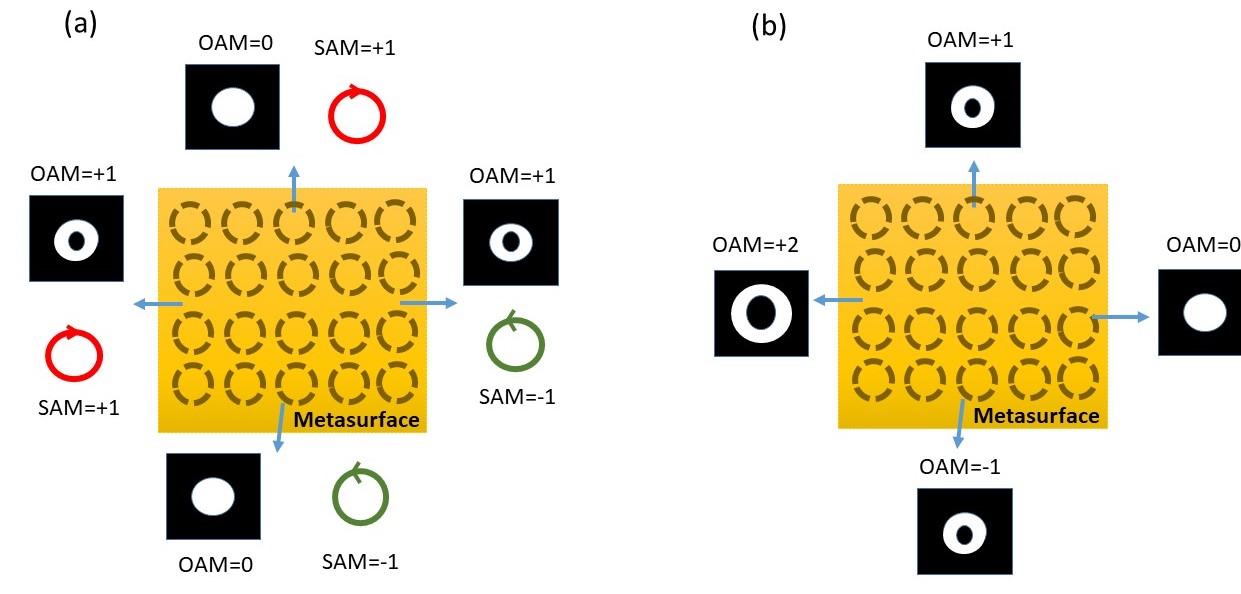}
\caption{OAM multiplexing techniques in metasurfaces.  (\textbf{a}) Schematic illustration of combined  SAM + OAM multiplexage, the multiplexed channels are based on a combination of polarization (SAM) and orbital angular momentum (OAM), the total angular momentum (OAM + SAM) spans a subspace of dimension $N=4$, (\textbf{b}) Schematic illustration of OAM multiplexing, the multiplexed channels are based orbital angular (OAM) momentum only, it spans a subspace {equivalent} dimension $N=4$. }
%\end{centering}
\label{fig5}
\end{figure}

Recently \cite{Vyas2000,Naidoo2000, Akihiko2000, Lin2000, Ngcobo2000},  an OAM multiplexing scheme in the Terahertz (THz) band based on use of a single-layer metasurfaces was demonstrated experimentally for the first time. The designed structure generates four focused phase vortex beams that have different topological charges (or OAM number~$l$) under illumination by a Gaussian beam, which means that OAM multiplexing with four channels is realized (see Figure \ref{fig5}). When an individual vortex beam is used as the incident light, only one channel is identified and extracted as a focal spot; that is, demultiplexing of the OAM signal is achieved. The~metasurface  structure has subwavelength-level thickness, which enriches the number of potential approaches available for the miniaturization and integration of THz communication systems. The~performance of the designed OAM multiplexing and demultiplexing device shows excellent agreement between the theoretical predictions and experimental results,  indicating that this device is suitable for scalable ultra high-speed THz communications.

\subsection{OAM Multicasting}

OAM multicasting consists of generating multiple coaxial OAMs from a single input,  a promissing  feature  for efficient optical  processing in one-to-many communications. This approach speeds up the end users to acquire the duplicate data, by replicating data into orthogonal multiple channels in the optical domain. Optical multicasting is used in many optical communication applications, such as interactive distance learning, video distribution, 
teleconferencing and distributed computing. \mbox{To this end}, Spatial Light Modulators (SLM) can be  loaded with suitably designed phase
patterns. A~pattern search assisted iterative (PSI) algorithm was proposed to {allow} simultaneous generation of multiple OAM modes using a single phase-only
SLM elements \cite{review1}. In addition, metasurface structures with a V-shaped antenna array can also be used to realize on-chip multicasting, from single Gaussian
beams to multiple OAM beams \cite{review1}. Additionally, adaptive multicasting assisted by feedback, 
from~a single Gaussian mode to multiple OAM modes was demonstrated using a complex phase pattern. Finally,~OAM multicasting by  manipulation  of
 amplitude and phase, turbulence compensation of distorted OAM multicasting by adaptive optics and data  carrying OAM-based underwater wireless optical multicasting link were also demonstrated, revealing the exceptional  performance of OAM multicasting and its broad applications  \cite{review1}. 

\subsection{Discussion}

{To a large extent}, the realization of spin-orbit-convertion (SOC) in metasurfaces and metamaterials may have a significant impact on diverse scientific and technological areas and can open a number of exciting research directions. The 2D nature of $q$-plates and generalized $J$-plates  enables {by and large} easy integration and holds the promise of solving scalability issues. Such 2D metasurfaces could {possibly} be integrated into laser cavities producing arbitrary Laguerre-Modes OAM states as well as tunable TAM states \cite{DelvinScience}. Additionally, we have described multiplexing and multicasting applications for encoding information in controllable optical states using both polarization (SAM) and azimuthal phase (OAM) which could {potentially} increase the bandwidth of optical communication beyond state of the art. In addition, controllable structured light produced by $J$-plates can, {to a great extent}, be used for the fabrication of new materials via laser imprinting, optical microengineering and~stimulated-emission-depletion imaging. {Although classical fields are mainly considered here, the physics of SOC can hold, to some extent, at the single-photon level. In particular, the transition to the quantum regime can be particularly promising in metasurfaces realization with demonstrated thigh transmission efficiency \cite{Faraon2015}}. For all these reasons, SOC in metamaterials and metasurfaces can {plausibly} furnish us with a new tool for both classical and  quantum communication, especially in schemes that rely  on polarization to OAM conversion.

\section{Conclusions}
\label{sec6}
2D metamaterials  and metasurfaces is a fast growing interdisciplinary field, with significant applications in nanophotonics, biophysics, plasmonics, quantum optics or~telecommunication. {To a large extent}, they warrant precise control of optical fields with unprecedented flexibility and performance. The reduced dimensionality of  optical metasurfaces enables applications that are distinctly different from those achievable with bulk metamaterials. {By~and large}, metasurfaces can provide for a novel tool for generation and conversion of OAM, a~feature which can promote many applications in integrated on-chip OAM multiplexing approaches, which~{conceivably} hold the promise of increasing transmission capacity and overcome scalability issues beyond state of the art. 

\vspace{6pt} 

\section{Acknowledgments}
GP wrote the entire manuscript and prepared the figures (both for version 1 and version 2 displayed on arXiv). Andrei Lavrinenko and Ricardo Depine suggested relevant references for version 1. Version 2 was accepted for publication in Quantum Reports 2019. The author is  grateful to Joerg Goette, Andrei Lavrinenko, Osamu Takayama, Ricardo Depine, and Juan Torres for helpful discussions. This work was supported by ANPCyT via grant PICT Startup 2015 0710 and UBACYT PDE 2017.
%\section{Acknowledgments}
%
%The author is  grateful to Joerg Goette, Andrei Lavrinenko, Ricardo Depine, and Juan Torres for helpful discussions. This work is supported by UBACyT PDE 2017 and PICT Startup 2015 0710.\\

 %Declare conflicts of interest or state ``The authors declare no conflict of interest.'' Authors must identify and declare any personal circumstances or interest that may be perceived as inappropriately influencing the representation or interpretation of reported research results. Any role of the funders in the design of the study; in the collection, analyses or interpretation of data; in the writing of the manuscript, or in the decision to publish the results must be declared in this section. If there is no role, please state ``The funders had no role in the design of the study; in the collection, analyses, or interpretation of data; in the writing of the manuscript, or in the decision to publish the results''.
%%%%%%%%%%%%%%%%%%%%%%%%%%%%%%%%%%%%%%%%%%

\end{document}